\documentclass{aa}
\usepackage{epsfig}
\def\be{\begin{equation}}
\def\ee{\end{equation}}
\begin{document}

\thesaurus{02     
               (02.07.2;  
                08.14.2)} 

\title{Cataclysmic Variables as Sources of Gravitational Waves}

\author{M. T. Meliani, J. C. N. de Araujo
and O. D. Aguiar}

\offprints{J.C.N. de Araujo}

\institute{Divis\~ao de Astrof{\'\i}sica, Instituto Nacional de
Pesquisas Espaciais, Avenida dos Astronautas 1758, \\ S\~ao Jos\'e
dos Campos, SP, 12227-010, Brazil. \\ E-mails:
meliani@das.inpe.br, jcarlos@das.inpe.br and odylio@das.inpe.br}

\date{Received ; accepted }

\titlerunning{Gravitational Waves -- Cataclysmic Variables}
\authorrunning{Meliani, de Araujo and Aguiar}
\maketitle

\begin{abstract}

General Relativity predicts that binary systems of stars produce
gravitational waves of significant intensity. Here we are
particularly interested in the cataclysmic variable binaries
(CVs). These systems emit low frequency gravitational waves, $ f <
10^{-3} Hz$. We present here a catalog of CVs and argue that part
of them are capable of being detected by the Laser Interferometer Space
Antenna (LISA).


\end{abstract}

\section{Introduction}

       Detection of gravitational radiation from astrophysical sources will
mark a breakthrough in the history of astronomy (see, e.g., Thorne
\cite{thor87} and Schutz \cite{schu96}). Experimental efforts to search
for these space-time wrinkles have been under
development for the past twenty years (Thorne \cite{thor95,thor96}).
With the advent of technological improvements in several crucial
aspects of the detection process we will soon be ready to turn them
a physical reality (Schutz \cite{schu96}, Thorne \cite{thor95},
Finn \&  Chernoff \cite{finn93}).

In particular, the Laser Interferometric Space Antenna (LISA) is
designed to detect low frequency gravitational waves in the
frequency range $10^{-4} - 1$ Hz, which are not possible to
detect on the Earth because of seismic noise. There is a lot of
very interesting astrophysical phenomena which are believed to
generate GWs in the frequency band detectable by LISA, namely:
formation of supermassive  black holes (SMBHs), SMBH-SMBH binary
coalescence, compact stars orbiting around SMBHs (in, e.g.,
galactic nuclei), a wide variety of binaries, such as pairs of
close white dwarfs (WDs), pairs of neutron stars, neutron star and
black hole binaries, pairs of contacting normal stars, normal
stars and white dwarfs (cataclysmic) binaries, and pairs of
stellar black holes.

Due to the fact the GWs are produced by a large variety of astrophysical
sources and cosmological phenomena it is quite probable that the Universe
is pervaded by a background of such waves. Binary stars of a variety
of stars (ordinary, compact or combinations of them), Population III
stars, phase transitions in the early Universe, cosmic strings are
examples of sources able to produce a background of GWs.

As the GWs possess a very weak interaction with matter passing
through it unharmed, relic radiation (spectral
properties for example) once detected can provide information on
the physical conditions from the era in which they were produced.
In principle it will be possible, for example, to get information
from the epoch when the galaxies and stars started to form and
evolve.

Concerning our galaxy, it presents a large number of binary systems,
which produce a GW background named binary confusion noise
(see Hils, Bender \& Webbink \cite{hils90}, Bender \& Hils
\cite{bend97}). Some of the galactic binary sources are: close
white dwarfs binaries (CWDBs), neutron star binaries (NSBs),
unevolved binaries, WUMs binaries and cataclysmic binaries.

The binary systems are the most understood of all sources of
GWs (see, e.g., Thorne 1987). Knowing  the masses of the stars, the
orbital parameters and their estimated distances, one can calculate the
details of the GW produced.

The LISA's sensitivity as well as the binary confusion noise will
determine in the end if one is able to discriminate the signal of a
particular astrophysical source.

The first papers concerning the gravitational radiation
from binaries systems was written by Mironowskii (1966), who studied in
particular the W UMa stars, and by Forward \& Berman (1967), approximately
30 years ago. After that many other studies concerning
the evaluation of GWs background produced by various
types of binary stars in the Galaxy followed (see, e.g., Douglass \&
Braginsky 1979, Lipunov \& Postnov
1987, Lipunov, Postnov \& Prokhorov 1987, Evans, Iben \& Smarr
1987, Hils, Bender \& Webbink 1990, Bender \& Hils 1997, Webbink \& Han 1998, Hils 1998)

Here we are particularly interested in the cataclysmic variable
binaries as sources of GWs, such a system is formed by a white
dwarf and a low mass secondary star. The total number of such a
kind of binary is estimated to amount $10^6$ in the Galaxy (see,
e.g. Hils, Bender \& Webbink 1990). These systems produce low
frequency GWs, namely, $f_{gw} < 10^{-3}$, which could be detected
by LISA.

We are not concerned here with the calculation of a confusion
noise produced by such binaries, our aim is similar to the study
by Douglass \& Braginsky (1979) who evaluate the dimensionless
amplitude h for a series of specific low frequency GW binaries.
Based mainly on the 6th edition of the catalogue of cataclysmic
binaries, low mass X-ray binaries and related objects (Ritter \&
Kolb 1998) we have catalogued almost 160 CV systems for which it is
possible to evaluate the GW amplitude. We have catalogued firstly those
CVs with known distances, orbital period and masses, quantities
necessary to evaluate the GW amplitude produced by such objects;
secondly we have catalogued those systems for which the distances and
the orbital periods are known, the masses being obtained from a
mass-period relationships.

The remainder of the paper is as follows: Section 2 deals with the
cataclysmic variables. Section 3 addresses the gravitational waves from
cataclysmic variables. The discussion and conclusions are summarized in
Section 4.

\section{The Cataclysmic Variables}

\begin{table*}
\caption[]{Catalogue of 68 CVs for which the distances, the
periods and the masses are known. In the columns we see, respectively, CV
names, distances in parsecs, periods in days, primary mass (in solar masses),
secondary mass (in solar masses), gravitational wave amplitude $h$ (see section 3
for its calculation), and references used to obtain the data of each CV system.}
\begin{flushleft}
\label{table 1a}
\begin{tabular}{lllllll}
            \hline
           \noalign{\smallskip} Name & d(pc) & P(days) &
$M_1/M_\odot$ & $M_2/M_\odot$ & log h & Ref. \\
            \noalign{\smallskip}
            \hline
            \noalign{\smallskip}

RX And      &      135 &    0.209893  &   1.14 &    0.48 & -21.16   & W87, RK98 \\
V603 Aql    &      110 &    0.1381    &   0.66 &    0.29 & -21.32 &  B96, RK98 \\
V1315 Aql   &     300  &   0.139690   &  0.73  &   0.30  & -21.72 &  RvPT92, RK98 \\
AE Aqr      &    102   &  0.411656    & 0.79  &   0.50 & -21.34 &  TK98, RK98 \\
HU Aqr      &    111   &  0.086820   &  0.95  &   0.15 & -21.35&   SHM96, RK98 \\
UU Aqr      &    200   &  0.163580   &  0.67  &   0.20 & -21.78&  BSH96, RK98 \\
T Aur       &    830   &  0.204378   &  0.68  &   0.63 & -22.02&   P84, RK98 \\
QZ Aur      &   2000   &  0.357496   &  1.05  &   1.05  & -22.22& CS95, RK98 \\
SS Aur      &    200   &  0.1828     &  1.08 &    0.39  & -21.39&  W87, RK98 \\
V363 Aur    &    600   &  0.321242    & 0.86 &    0.77  & -21.85&  RvPT92, RK98 \\
Z Cam       &    175   &  0.289841   &  0.99  &   0.70 & -21.27&   W87, RK98 \\
OY Car      &     86   &  0.63121    &  0.685  &  0.070 & -22.23& BBB96, RK98 \\
HT Cas      &    165   &  0.073647   &  0.61  &   0.09 & -21.82&   W87, RK98 \\
BV Cen      &    450   &  0.610108   &  0.83   &  0.90   & -21.87& P84, RK98 \\
V436 Cen    &    210   &  0.062501   &  0.7   &   0.17  & -21.57&  W87, RK98 \\
WW Cet      &    100   &  0.1758     &  0.85   &  0.41   & -21.14& W87, RK98 \\
Z Cha       &    130   &  0.074499   &  0.84  &   0.125 & -21.49&  W87, RK98 \\
HL CMa      &    210   &  0.2145     &  1.0    &  0.45  & -21.43&  W87, RK98 \\
BG CMi      &    700   &  0.134749   &  0.8   &   0.38 & -21.96&   W95, RK98 \\
AC Cnc      &    800   &  0.300478   &  0.82    & 1.02   & -21.87& W87, RK98 \\
SY Cnc      &    450   &  0.380      &  0.89 &    1.10  & -21.63&  W87, RK98 \\
YZ Cnc      &    290   &  0.0868     &  0.82   &  0.17   & -21.76& W87, RK98 \\
TV Col      &    500   &  0.228599   &  0.75  &   0.56  & -21.84&  W95, RK98 \\
TX Col      &    550   &  0.2383     &  1.3    &  0.57   & -21.70& W95, RK98 \\
TV Crv      &    350   &  0.06250    &  0.52   &  0.12 & -22.03&  HRAH96 \\
EM Cyg      &    350   &  0.290909   &  0.57   &  0.76   & -21.74& W87, RK98 \\
SS Cyg      &     75   &  0.275130   &  1.19   &  0.704  & -20.82& W87, RK98 \\
CM Del      &    280   &  0.162      &  0.48   &  0.36   & -21.81& W87, RK98 \\
HR Del      &    285   &  0.214165   &  0.67   &  0.55  & -21.62&  W87, RK98 \\
DO Dra$^{\rm a}$ & 155 &  0.165374   &  0.83   &  0.38   & -21.35& W95, RK98 \\
EP Dra      &    300   &  0.072656   &  0.43   &  0.13   & -22.04& W95, RK98 \\
U Gem       &     81   &  0.179606   &  1.26    & 0.57   & -20.79& W87, RK98 \\
AH Her      &    250   &  0.258116   &  0.95   &  0.76   & -21.37& W87, RK98 \\
AM Her      &     75   &  0.128927   &  0.39    & 0.26   & -21.36& W95, RK98  \\
DQ Her      &    330   &  0.193621   &  0.60   &  0.40   & -21.81& W87, RK98 \\
V838 Her    &   3000   &  0,2976635  &  0.87    & 0.74  &  -22.54&VSWS96, RK98 \\
EX Hya      &    105   &  0.068234   &  0.78  &   0.13  &  -21.37& W95, RK98 \\
VW Hyi      &     65   &  0.074271   &  0.63   &  0.11   &  -21.33&W87, RK98 \\
WX Hyi      &    265   &  0.074813   &  0.90  &   0.16  &  -21.67& W87, RK98 \\
DP Leo      &    450   &  0.062363   &  0.71   &  0.11   & -22.08& W95, RK98 \\
T Leo       &     76   &  0.05882    &  0.16   &  0.11   & -21.77& SHM96, RK98 \\
ST LMi      &    128   &  0.079089   &  0.76   &  0.17   & -21.40& W95, RK98 \\
BT Mon      &   1700   &  0.333814   &  1.04  &   0.87 &   -22.20&SDM98, RK98 \\
V426 Oph    &    100   &  0.2853     &  0.90   &  0.70   &  -21.05&W87, RK98 \\
V2951 Oph   &    140   &  0.062428   &  0.44   &  0.13   &  -21.65&P84, RK98 \\
CN Ori      &    295   &  0.163199   &  0.74   &  0.49   &  -21.57&W87, RK98 \\
EF Peg      &    172   &  0.0837     &  0.65   & 0.17   &  -21.59&SHM96, RK98 \\
IP Peg      &    124   &  0.158206   &  1.15   &  0.67   &  -20.91&W87, RK98 \\
RU Peg      &    174   & 0.3746      & 1.21    & 0.94   &  -21.16&W87, RK98 \\
GK Per      &    340   &  1.996803   &  0.90   &  0.25   & -22.55& W95, RK98 \\
SW Sex$^{\rm b}$ & 450 &  0.134938   &  0.58   &  0.33    & -21.93&RvPT92, RK98 \\

           \noalign{\smallskip}
            \hline
\end{tabular}
\end{flushleft}
\begin{list}{}{}
\item[$^{\rm a}$] Also known as YY Dra
\item[$^{\rm b}$] Also known as PG 1012-03
\end{list}
\end{table*}
\setcounter{table}{0}
\begin{table*}
\begin{flushleft}
\caption[]{\bf\it Continued}
\label{table 1b}
\begin{tabular}{lllllll}
            \hline
           \noalign{\smallskip} Name & d(pc) & P(days) &
$M_1/M_\odot$ & $M_2/M_\odot$ & log h & Ref. \\
            \noalign{\smallskip}
            \hline
            \noalign{\smallskip}

RR Pic      & 240   & 0.145025  &  0.95  &  0.4  & -21.43 & B96, RK98 \\
VZ Scl      & 530   &  0.144622 &  1.0   &  0.4  & -21.76 & W87, RK98 \\
LX Ser      & 340   &  0.158432 &  0.41  &  0.36 & -21.94 & RvPT92, RK98 \\
RW Sex      &    290   &  0.24507    &  0.8    &  0.6& -21.57    &  W87, RK98 \\
V Sge       &     56   &  0.514197   &  0.74   &  2.8& -20.57   & B96, RK98 \\
WZ Sge$^{\rm a}$ &194  &  0.056688   &  0.45   &  0.058&-22.09  & B96, RK98 \\
V1223 Sgr   &    600   &  0.140244   &  0.5     & 0.4  &-22.04   & W95, RK98 \\
V3885 Sgr   &    280   &  0.2163     &  0.8   &   0.7 & -21.46  &  W87, RK98 \\
RW Tri      &    270   &  0.231883   &  0.45   &  0.63& -21.72   & RvPT92, RK98 \\
SW UMa      &    140   &  0.05618    &  0.71   &  0.10& -21.58   & W87, RK98 \\
UX UMa      &    250   &  0.196671   &  0.47    & 0.47 & -21.72  & RvPT92, RK98 \\
CU Vel      &    200   &  0.0785     &  1.23   &  0.15 & -21.49 &  W87, RK98 \\
IX Vel$^{\rm b}$ & 150 &  0.193929   &  0.82    & 0.53 & -21.26   & W87, RK98 \\
TW Vir      &    455   &  0.18267    &  0.91   &  0.40 & -21.79 &  W87, RK98 \\
J1015.5+0904 &   100   &  0.054777 &  0.56-1.12 & 0.09 & -21.54  &BRSH98 \\
DX And & 660 & 0.440502 & 0.51 & 0.50 & -22.33    &SD98, SHM96, RK98 \\
VV Pup & 145 &  0.69749 & 1.1 & 0.2 &  -21.90     &RK98, KB99 \\
           \noalign{\smallskip}
            \hline
\end{tabular}
\end{flushleft}
\begin{list}{}{}
\item[$^{\rm a}$] Smack (1993) obtained a distance of $48 pc$, which gives
${\rm log\, h = -21.49}$, a factor of 4 greater.
\item[$^{\rm b}$] Not always classified as CV
\end{list}
\end{table*}

\begin{table*}
\caption[]{Catalogue of 88 CVs for which only the distances and
the periods are known. In the columns we see, respectively, CV
names, distances in parsecs, periods in days, gravitational wave
amplitude $h$ (see section 3 for its calculation), and references
used to obtain the data of each CV system.}
\begin{flushleft}
\label{table 2a}
\begin{tabular}{lllll}
            \hline \noalign{\smallskip} Name & d(pc) & P(days)&
            log h & Ref.
\\
            \noalign{\smallskip}
            \hline
            \noalign{\smallskip}

AR And     &     269 &    0.1630     & -21.59 &   SHM96, VBRP97,RK98 \\
DH Aql     &     116 &    0.0778     & -21.51&     SHM96 ,RK98 \\
UU Aql     &     225 &    0.14049    & -21.55&     SHM96 , RK98 \\
V1101 Aql  &     300 &    0.1441667  & -21.67&      MdV98 \\
V1432 Aql$^{\rm a}$  & 230 & 0.140235 & -21.57&    W95, RK98 \\
FO Aqr     &     325 &    0.202060   & -21.63&     W95, RK98 \\
VY Aqr     &      97  &    0.0635     & -21.55&     SHM96, RK98 \\
TT Ari     &     185  &   0.137551    & -21.47&     W87, RK98 \\
XY Ari     &     200  &   0.2526697   & -21.39&     W95, RK98 \\
WX Ari     &     198  &   0.13934     & -21.50&     SHM96, RK98 \\
RS Cae     &     440  &   0.07        & -22.14&     BRSB96\\
AF Cam     &     425  &   0.23        & -21.73&     SHM96, RK98 \\
BY Cam     &     190  &   0.13979     & -21.48&   W95, RK98,DM98 \\
BZ Cam     &     830  &   0.153693    & -22.09&      RN98, RK98 \\
QU Car     &     500  &   0.454       & -21.72&     W87, RK98 \\
V592 Cas   &     330  &   0.115063    & -21.79&      TTPF98 \\
V705 Cas   &    2400  &   0.2280      & -22.48&     MGWS98, RK98 \\
V834 Cen   &      86  &   0.070498    & -21.43&      W95, RK98 \\
WX Cet     &     185  &   0.05829     & -21.89&     SHM96, RK98 \\
AR Cnc     &     681  &   0.2146      & -21.95&     SHM96, RK98 \\
EG Cnc     &     320  &   0.05997     & -22.11&      PKS98\\
UU Col     &     740  &   0.143750    & -22.06&       BRBT96\\
AL Com     &     190  &   0.056668    & -21.93&       SHM96, RK98 \\
GO Com     &     361  &   0.0658      & -22.10&     SHM96, RK98 \\
GP Com     &      90  &   0.03231     & -22.22&     P84 \\
V394 CrA   &    5000  &   0.7577      & -22.69&   W95, RK98 \\
V1500 Cyg  &    1200  &   0.139513    & -22.28&   W95, RK98 \\
V1521 Cyg  &   10000  &   0.1997      & -23.12&    P84 \\
V1668 Cyg  &    3600  &   0.1384      & -22.76&   P84, RK98 \\
V1974 Cyg  &    1770  &   0.081259    & -22.67& CGPK97, RK98 \\
DM Dra     &     580  &  0.087       &  -22.13& SHM96, RK98 \\
CQ DraBC   &     100  &   0.1256      & -21.23&  RGB98, RK98 \\
AH Eri     &     113  &   0.2391      & -21.15&  SHM96, T97\\
EF Eri     &      94  &   0.056266    & -21.63&  W95, RK98 \\
UZ For     &     230  &   0.087865    & -21.73&  W95, RK98, SMB97 \\
IR Gem     &     250  &   0.0684      & -21.91&     W87, RK98 \\
V533 Her   &    1200  &   0.2098      & -22.20&      P84, RK98 \\
WW Hor     &     430  &   0.080199    & -22.06&    W95, RK98 \\
BL Hyi     &     128  &   0.078915    & -21.55&  W95, RK98 \\
DO Leo     &     878  &   0.234515    & -22.04& SHM96, RK98 \\
RZ Leo     &     174  &   0.0708      & -21.73& SHM96, RK98 \\
X Leo      &     345  &   0.1644      & -21.70& W87, RK98 \\
RU LMi     &    1273  &   0.251       & -22.19& SHM96, RK98 \\
SX LMi     &     150  &   0.0625      & -21.75& SHM96, RK98 \\
BK Lyn     &     114  &   0.07498     & -21.52& SHM96, RK98 \\
AY Lyr     &      52  &   0.07370     & -21.19& SHM96, RK98 \\
CY Lyr     &      115 &   0.1591      & -21.23& W87, TTK98 \\
MV Lyr     &     322  &   0.1329      & -21.72&  W87, RK98 \\
TU Men     &     270  &   0.1172      & -21.70& W87, RK98 \\
CW Mon     &     290  &   0.1762      & -21.61& W87, RK98 \\
CQ Mus     &     290  &   0.059365    & -22.07& VBRP97, RK98 \\

            \noalign{\smallskip}
            \hline
\end{tabular}
\end{flushleft}
\begin{list}{}{}
\item[$^{\rm a}$] Also known as J1940.2-1025
\end{list}
\end{table*}

\setcounter{table}{1}

\begin{table*}
\caption[]{\bf\it Continued}
\begin{flushleft}
\label{table 2b}
\begin{tabular}{lllll}
            \hline \noalign{\smallskip} Name & d(pc) & P(days) &
            log h & Ref.
\\
            \noalign{\smallskip}
            \hline
            \noalign{\smallskip}
V841 Oph   &     255  &   0.60423     & -21.41&  W87, RK98 \\
CZ Ori     &     300  &   0.2189      & -21.59&  W87, RK98 \\
V1309 Ori$^{\rm a}$ &1500 & 0.332613& -22.23  &  HCPD97, RK98 \\
V349 Pav$^{\rm b}$  & 400 & 0.1109 & -21.89   &   W95, RK98 \\
KT Per     &     245  &   0.162500 & -21.55  &   TR97 \\
TZ Per     &     275  &   0.2605  & -21.52   &   W87, RK98 \\
UV Per     &     115  &   0.0622  & -21.64   &  W87, RK98 \\
TY PsA     &     190  &   0.0841  & -21.66   &  W87, RK98 \\
AO Psc     &     420  &   0.149626& -21.81   &  W95, RK98 \\
AY Psc     &     565  &   0.217321 & -21.86  &  SHM96, RK98 \\
TY Psc     &     250  &   0.6833  & -21.39   &  W87, RK98 \\
BX Pup     &     750  &   0.127   & -22.10   &  W87, RK98 \\
CP Pup     &     556  &   0.06143 & -22.33   &  B96, RK98 \\
U Sco      &   14000  &  1.23056  & -23.11   &   W95 \\
MR Ser     &     139  &   0.78798 & -21.13   &  W95, RK98 \\
UZ Ser     &     300  &   0.1730  & -21.63   &  W87, RK98 \\
WY Sge     &     700  &   0.153635& -22.02   &  SMN96, RK98 \\
QS Tel     &     300  &   0.097187 & -21.81  &  W95, RK98 \\
EK Tra     &     200  &   0.0636  & -21.86   &  W87, RK98 \\
AN UMa     &     270  &   0.79753 & -21.41   & W95, RK98, BMSS96 \\
BC UMa     &     255  &   0.063  & -21.97    &  SHM96, RK98 \\
DI UMa     &     107  &   0.0548 & -21.71    &   SHM96, RK98 \\
DV UMa$^{\rm c}$ & 277&   0.08597& -21.82    &   SHM96, RK98  \\
EV UMa     &     700  &   0.055338& -22.52   &   W95, RK98 \\
SU UMa     &     280  &   0.7635  & -21.43   &   W87, RK98 \\
PW Vul$^{\rm d}$ & 1600 &   0.2137& -22.32   &  RN96, RK98  \\
QQ Vul     &     320  &   0.154520& -21.68   &   W95, RK98 \\
QU Vul     &    2600  &   0.111765 & -22.70  &  dVGB97, RK98 \\
E2259+586  &    3600  &   0.0266  &  -23.77  &    P84 \\
J0132.7-6554 &   300  &   0.0540499& -22.17  &   BRBT97 \\
J0203.8+2959 &   600  &   0.191667& -21.91   &   SSB98, RK98 \\
J0744-52     &   820  &   0.15   &  -22.10   &   RBC98 \\
J1016.9-4103 &   615  &   0.093055& -22.13   &   GS98 \\
J1724.0+4114 &   250  &   0.0832639& -21.81  &   GSW98 \\
J1957.1-5738$^{\rm e}$ & 350 & 0.068625& -22.06  &TBSB96, RK98 \\
J2022.6-3954 &   190  &   0.05417889&  -21.97& BRBT97 \\
J2115.7-5840$^{\rm f}$ & 250 & 0.07691 & -21.85 & SBOH97, RK98 \\

            \noalign{\smallskip}
            \hline
\end{tabular}
\end{flushleft}
\begin{list}{}{}
\item[$^{\rm a}$] Also known as J0515.6+0105
\item[$^{\rm b}$] Also known as V2008-65.5
\item[$^{\rm c}$] Not always classified as CV
\item[$^{\rm d}$] Not always classified as CV
\item[$^{\rm e}$] Also known as Pav4
\item[$^{\rm f}$] Also known as Ind1
\end{list}
\end{table*}
%

A Cataclysmic Variable (CV) is a semi-detached binary system of
low mass and very short orbital period. The primary star is an
accreting degenerate white dwarf and the secondary one is usually,
but not always, a late-type star that fills its critical Roche
lobe and transfers matter to the companion. There are 1020
cataclysmic variables classified (Downes, Webbink \& Shara, 1997)
and more than 300 of them have known periods (Ritter \& Kolb,
1998, hereafter RK98). From a period histogram Patterson (1998,
his Figure 1) shows, with data taken from RK98 (see also Kolb,
King and Ritter, 1998, figure 4, to orbital periods below 5
hours), that the majority of these systems have periods ranging
from 1.2 to 15.0 h.

    We have catalogued, in Tables 1 and 2, 156 CV systems. In
Table 1 we have catalogued 68 CVs, where in column 1 we present their
names, in column 2 the distances in parsecs, in column 3 the
periods in days, in column 4 the primary mass in solar masses, in
column 5 the secondary mass in solar masses, in column 6 the
gravitational wave amplitude $h$ (see section 3 for its
calculation), and finally in column 7 we present the references
used to obtain the data of each CV system. In Table 2 we have catalogued
88 CVs for which only the distances and periods are known; the
label of the columns are the same as in  Table 1.

For the systems with orbital periods of up to 10 hours it is
possible to make use of a mass function to compute the masses of
the secondary stars. We have computed the mass of the secondary
star using an equation obtained by Smith and Dhillon (1998,
hereafter SD98). Their  mass-period relationship has the following
best fit (equation 9 of SD98):

\begin{equation}
M_2/M_\odot = 0.126P- 0.11, \;\; {\rm with\; period\; in\; hours.}
\end{equation}

To calculate the mass of the primary star we have used the
unweighted average for all systems (see, Table 4a of SD98):

\begin{eqnarray}
M_1= 0.69 M_\odot  \;\; {\rm below\; period\; gap}\nonumber\\ M_1=
0.80 M_\odot  \;\; {\rm above\; period\; gap}
\end{eqnarray}

The period gap, namely, $ 2 < P < 3 $ hours, a failure in the
distribution of cataclysmic variables, has been discussed in the
literature recently by, for example, Clemens et al.(1998) and
Kolb, King and Ritter (1998). For stars in the gap period we have
considered the mean value $M_1=0.74 M_\odot$ .

It is worth noting that Equations 1 and 2 (SD98) were obtained
from 14 reliable CV mass determination. In our sample there are 68
CVs with known masses, whose values were obtained by various
methods. A fit with 62 CVs gives a relationship consistent with
SD98. Five do not fit the $M_2$ $\times$ orbital period
distribution, namely: AE Aqr, OY Car, BV Cen, GK Per, V Sge. Our
fit is given by:

\begin{eqnarray}
M_2/M_\odot = (0.121 \pm 0.004)P - 0.070 \pm 0.020,
\\ {\rm with\; period\; in\; hours.
\qquad\qquad\qquad\qquad\qquad\;}\nonumber
\end{eqnarray}

In our catalogue 9 systems have periods above 9 hours, namely: QU
Car, V394 CrA, V841 Oph, TY PsC, VV Pup, U Sco, MR Ser, NA UMa and
SU UMa. From RK98 we have obtained the spectral type only for VV
Pup (M4-5), U Sco (F6-G0-5) and MR Ser (M5-6/5). The secondary
mass is then obtained from the spectral type versus $M_2$ diagram of
Kolb \& Baraffe (1999). For VV Pup RK98 give a mass ratio $M_1/M_2=
5.5$, giving in this way $M_2= 0.2 M_\odot$ and $M_1= 1.1
M_\odot$; for U Sco $1.0<M_2<1.3 M_\odot$; and for MR Ser $M_2< 0.1
M_\odot$. For all these systems with the exception of VV Pup, we have
also to make use of the mass function. We have considered these
values as upper limits to the secondary mass.

\section{Gravitational Waves from Cataclysmic Variables}

We proceed now calculating the gravitational wave amplitude ($h$)
and frequency ($f_{gw}$) for the CVs presented in our catalogue.
As already mentioned the binary systems are the most understood of
all sources of GWs (see, e.g., Thorne 1987). Knowing the masses of
the stars, the orbital parameters and their estimated distances,
one can calculate the details of the GW produced. In our catalogue
for 68 of the CVs the necessary parameters for the calculation of
$h$ and $f_{gw}$ are known, for the other 88 CVs we needed to
obtain their masses through the equations 1 and 2, as discussed
in preceding section.

\begin{figure*}
\begin{flushleft}
\leavevmode
\centerline{\epsfig{figure=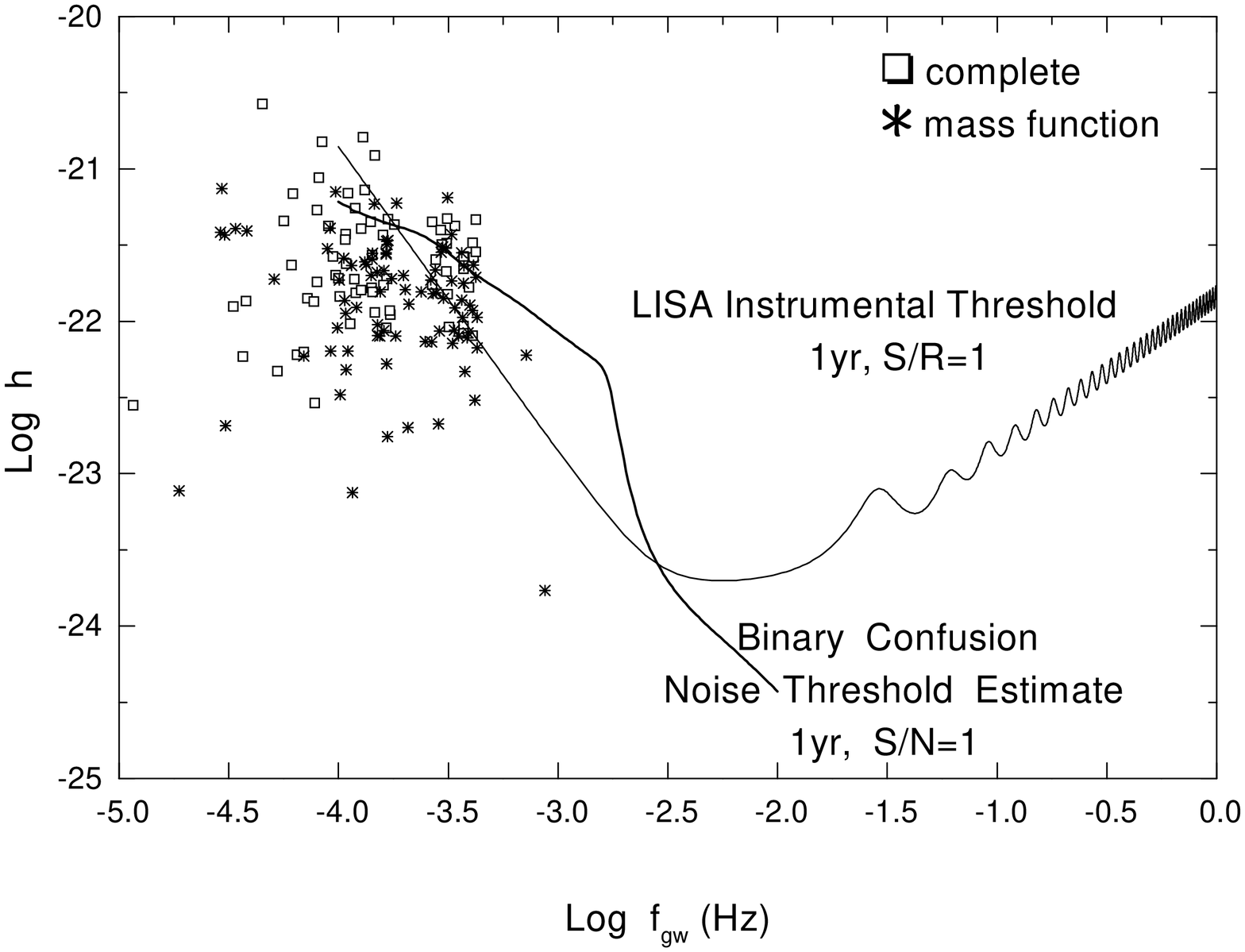,angle=360,height=15.0cm,width=20.0cm}}
\caption{Dimensionless amplitude h versus GW frequency $f_{gw}$ of
all CVs of our catalogue and the LISA instrumental threshold and
the binary confusion noise threshold estimate curves for 1 year of
observations and S/N=1.}
\end{flushleft}
\label{figall}
\end{figure*}

The CVs emit GWs at twice the orbital frequency and harmonics
thereof (see, e.g., Thorne 1987).  For eccentricity $\epsilon <
0.2$ the line at $f_{gw}=2f_{orb}$ is the dominant; for $\epsilon
\simeq 0.5$ the lines at $f_{gw}/f_{orb}\simeq 2$ through 8 are
all strong; for $\epsilon \simeq 0.7$ the lines at
$f_{gw}/f_{orb}\simeq 4$ through 20 are all strong (see, e.g.,
Thorne 1987). Following Thorne (1987), the characteristic
amplitude, in the low eccentricity case with $f_{gw}=2f_{orb}$, is
given by

\begin{eqnarray}
\lefteqn {h \, = \, 8.7 \, \times \, 10^{-21} \times {} }
\nonumber\\
& & {}\times
{\biggl( {\mu \over{M_{\odot}}} \biggr)}
{\biggl( {M \over{M_{\odot}}} \biggr)}^{2/3}
{\biggl( {100 \, pc \over{r}} \biggr)}
{\biggl({f \over 10^{-3} \, Hz} \biggr)}^{2/3}
\end{eqnarray}


\begin{figure*}
\begin{flushleft}
\leavevmode
\centerline{\epsfig{figure=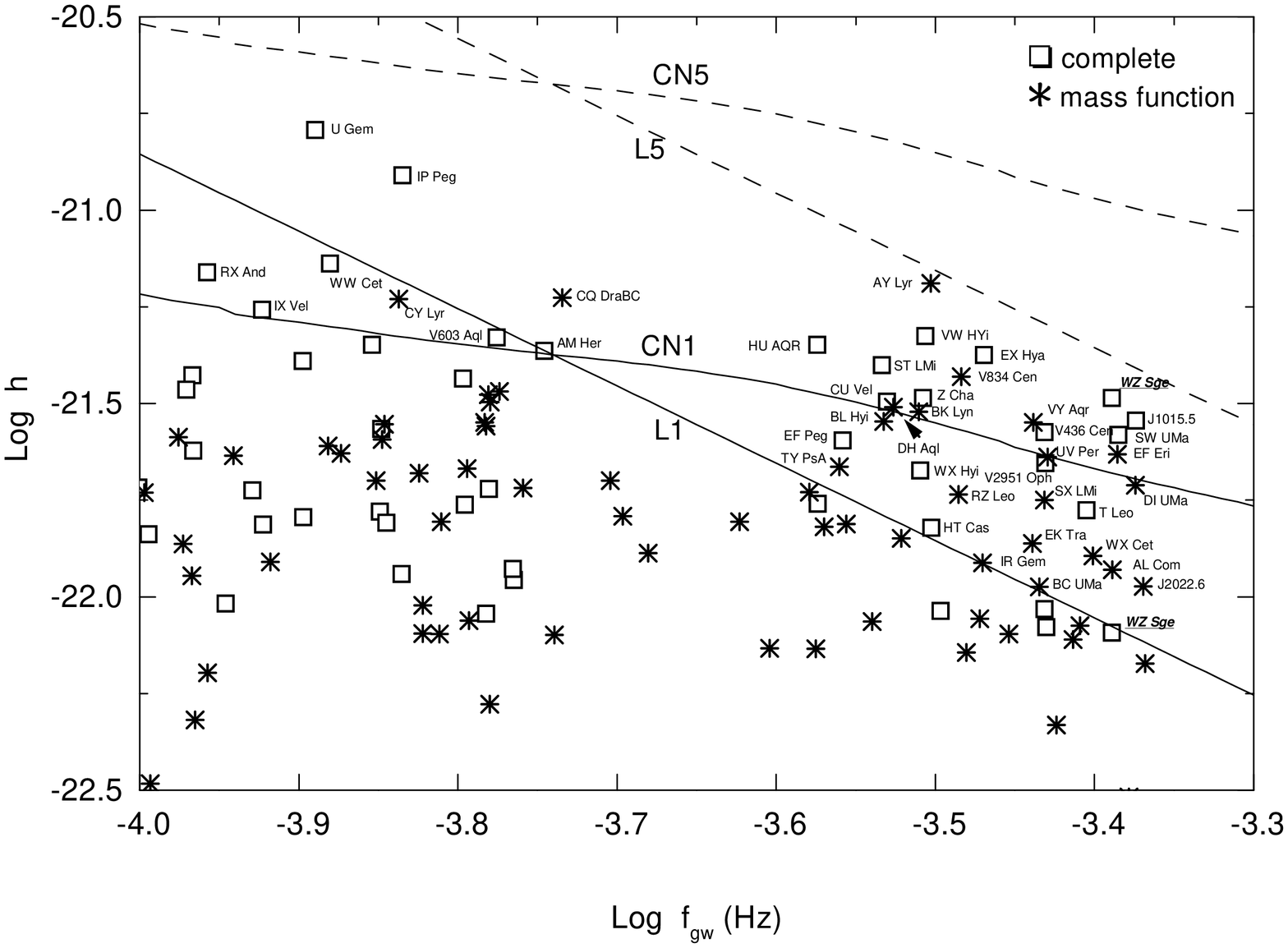,angle=360,height=15.0cm,width=20.0cm}}
\caption{Dimensionless amplitude h versus GW frequency $f_{gw}$
for some CVs of our catalogue. The curves labeled L1 (L5) and CN1
(CN5) are the LISA instrumental threshold and the binary confusion
noise threshold estimate curves for 1 year of observations and
S/N=1 (S/N=5), respectively.}

\end{flushleft}
\label{figL5}
\end{figure*}

The above equation takes into account both polarizations, $h_{+}$
and $h_{\times}$, and it is averaged over the orientation angles
of the source (Thorne 1987). The amplitude given by this equation
is thus a factor of  $\simeq 2$ smaller than the maximum
amplitude.

We are also considering that all CVs of our catalogue have low
eccentricity, and therefore equation 4 can be applied to them.

The LISA curves, as discussed by the LISA Study Team (1998) are
calculated realistically, and in some sense somewhat conservative,
due to the fact that the sensitivity could in principle be
improved in many aspects.

The LISA mission is planned to last 2 years, but it could last up to
10 years, as a result: a) its sensitivity to long-lived sources is
improved; b) the noise, the threshold curves and the GW noise from
white-dwarf binaries would lower, as a result it would be possible
to resolve more sources and remove them from the binary confusion
noise background.

Although the three LISA arms are not independent, LISA could in
some sense act as two interferometers, improving its capability of
detection and sensitivity. A third arm allows LISA to detect two
different GW observable, which can be thought of as being formed from
the signals of two different interferometers, with one arm common
to both. As a result, besides an improvement in sensitivity, LISA's
ability to measure, for example, the polarization of the GWs is
improved. It is worth mentioning that the LISA curves usually
presented elsewhere only consider a single interferometer.

In Figure 1 we show the dimensionless amplitude h (using equation
3), for all the CVs presented in Tables 1 and 2, as a function of
the GW frequency; also plotted are the curves for the LISA
instrumental threshold and the binary confusion noise threshold
estimate curves for 1 year of observations and S/N=1. The values
for h for all CVs of our catalogue, calculated via equation 4, are
also presented in Tables 1 and 2.

In Figure 2 we zoom Figure 1 for the frequency band
$1-5\times10^{-4}$ Hz, and also plot the curves labeled L1 (L5)
and CN1 (CN5) which are the LISA instrumental threshold and the
binary confusion noise threshold estimate curves for 1 year of
observations and S/N=1 (S/N=5), respectively.

Among the CVs presented in our catalogue no one has $S/N > 5$, and
therefore at this signal-to-noise ratio it is not possible to
detected them.

It is worth mentioning at this point that even the CV named WZ
Sge, which is usually considered to be one of the most promising
CVs capable of being detected by LISA, cannot be detected at
$S/N > 5 $. We have used here new data presented  mainly in the
6th edition of the catalogue of cataclysmic binaries, low mass
X-ray binaries and related objects (Ritter \& Kolb 1998), and in
particular for the WZ Sge the masses presented are smaller than
thought before (see, e.g., Douglass \& Braginsky 1979). This
explain why WZ Sge appears here in our study with a dimensionless
amplitude $h$ much smaller than presented by the LISA Study Team
(1998).

The parameters for WZ Sge used by Douglass \& Braginsky (1979)
were obtained from Warner (1976), namely, $M_1= 1.5 M_\odot$ and
$M_2= 0.12 M_\odot$ (the masses), and from Kraft (1962), namely,
$d= 75pc$ (the distance). Barret (1996), on the other hand,
obtained a distance of 194 pc (this is the distance that appears
in Table 1) with the linear polarimetric technique. Smack (1993),
instead, obtained the masses $M_1= 0.45 M_\odot$, $M_2= 0.058
M_\odot$ and a distance of $d= 48 pc$, from visual and ultraviolet
observations.  Even considering a distance of $d= 48 pc$,
WZ Sge would appear below $S/R=5$ curves. For comparison we plot
WZ Sge for a distance of $ d=48pc$ (see, Figure 2).

As usual in astrophysics the distance plays a key role here. The
case for WZ Sge is an example that we have addressed to call
attention to an issue that could occur with almost all other CVs
of our catalogue. As a result this uncertainty in the distance could
move the points plotted in Figures 1 and 2 upwards or downwards.

From our sample we note that 37 CVs  have $h$ values greater than
the S/N =1 LISA curve and also appear above the binary confusion
noise curve, such CVs, therefore, could in principle be detected
at this signal-to-noise ratio; Of these 37 CVs, 33 are below the
period gap ($1.25 < P < 2.16$ hours). We also note that the
maximum distance of these CVs to the Earth is approximately 300
pc. Patterson (1998) estimates that the space density of active
CVs is $d=10^{-5}pc^{-3}$, with 75\% of them below the period gap.
So, the  expected number of active CVs up to a radius of 300 pc
would amount to approximately 850 systems with periods below the
gap. We have therefore only a part of them in our catalogue. This
means that the prospect of detection of CVs is improved.

It is worth mentioning that even if the sources are not detectable
after 1 year of observation they can be detected after an additional
integration time t, namely

\begin{equation}
h_{\rm CV} > (f_{gw}\cdot t)^{-1/2}\; h_{\rm confusion\; noise}
\end{equation}

\par\noindent (see, e.g., Thorne 1987). It is important to have in mind,
however, that below 1 mHz there could exist many binaries per
frequency bin that could be hard to resolve individual sources
(see, e.g., Hils 1998), unless we know their position in sky like those
in Tables 1 and 2.

It is worth mentioning also that due to the fact that the LISA
curves presented here are only for single interferometers, and
that LISA could work as two independent interferometers, this
improves the possibilities of detection of CVs by LISA, since LISA
curves as well as the binary confusion noise curves go down.

\section{Discussion and conclusions}

The CVs produce GWs which could in principle be detected by the
LISA antenna, since CVs produce low frequency GWs in the frequency
band where LISA is sensitive. Due to the fact that a positive
detection of a CV by the LISA antenna might be improved once we
know the sources beforehand we compile in the present study a
catalogue of CVs, for which we know at least their orbital periods
and distances.

We argue that the present study is of interest since in the
literature one has not found a systematic identification of
possible detectable GW CVs, since an early study made by Douglass
\& Braginsky (1979) twenty years ago, and also a preliminary study
by Aguiar et al. (1998). We have been able to catalogue
approximately 160 CVs, from which a reasonable part of them could be
detected once the LISA antenna become operative.

We argue that it would be of interest whether other groups
performed a similar study for the other binary systems which
produce low frequency GWs in the frequency band where the LISA
antenna is sensitive.

It is worth mentioning that a positive detection of a binary
system through its gravitational emission, with some help of
electromagnetic data observations, could lead one to know all the
parameters related to the binary system, namely, the masses of the
stars, their distances to the earth, the period of the system and
their orientation angles.

\begin{acknowledgements}
MTM and JCNA thank FAPESP (Brazil) for financial support (grants
97/13415-0, 98/07641-0 and 97/06024-4, 97/13720-7, respectively).
ODA thanks CNPq (Brazil) for financial support (grant
300619/92-8). We would like to thank Dr. Robin Stebbins and Prof.
Peter Bender for kindly providing us with the LISA sensitivity
curve.
\end{acknowledgements}

\end{document}